\documentclass[12pt,preprint]{aastex}

\begin{document}
\title{$r$-Process Nucleosynthesis in Shocked Surface Layers
of O-Ne-Mg Cores}
\author{H. Ning and Y.-Z. Qian}
\affil{School of Physics and Astronomy, University of
Minnesota, Minneapolis, MN 55455}
\email{hning@physics.umn.edu; qian@physics.umn.edu}
\and
\author{B. S. Meyer}
\affil{Department of Physics and Astronomy, Clemson University,
Clemson, SC 29634}
\email{mbradle@clemson.edu}

\begin{abstract}
We demonstrate that rapid expansion of the shocked surface layers of
an O-Ne-Mg core following its collapse can
result in $r$-process nucleosynthesis.  As the supernova shock
accelerates through these layers, it makes them expand so
rapidly that free nucleons remain in disequilibrium with $\alpha$-particles 
throughout most of the expansion.  This allows heavy
$r$-process isotopes including the actinides
to form in spite of the very low initial neutron excess
of the matter.  We estimate that yields of heavy $r$-process nuclei
from this site may be sufficient to explain the Galactic inventory
of these isotopes.
\end{abstract}


\section{Introduction}

In this Letter we propose a new astrophysical site for $r$-process
nucleosynthesis. In particular, we show that heavy nuclei with mass
numbers $A>130$ through the actinides may be produced in the shocked
surface layers of an O-Ne-Mg core following its gravitational
collapse. Stellar evolution models show that stars of $\sim
8$--$11\,M_\odot$ develop degenerate O-Ne-Mg cores, at least some of
which eventually collapse to produce supernovae (SNe; e.g.,
\citealp{1984ApJ...277..791N, 1987ApJ...322..206N,
1994ApJ...434..306G, 1996ApJ...460..489R, 1999ApJ...515..381R,
1997ApJ...485..765G, 1997ApJ...489..772I, 2007arXiv0705.4643P}).
Such a core is separated from the hydrogen envelope by a thin C-O
shell and an even thinner He shell. However, the density falls off
so steeply in the region immediately above the core that we shall
refer to the C-O and He shells as the surface layers of the core.
These layers result from He and H burning coupled with convection
during the pre-SN evolution, and this pre-supernova burning can give
rise to a metallicity-independent neutron excess through reaction
sequences such as $^{12}{\rm C}(p,\gamma)^{13}$N followed by
$\beta^+$ decay of $^{13}$N to $^{13}$C (Nomoto 2007, personal
communication). The SN shock rapidly accelerates as it propagates
through the surface layers of the core. This gives rise to fast
expansion of the shocked ejecta on timescales of $\sim 10^{-4}$~s.
Together with an entropy of $S\sim 100$ (in units of Boltzmann's
constant $k$ per nucleon) and an initial electron fraction of
$Y_e\sim 0.495$ (e.g., for a composition of ${^{13}{\rm
C}}:{^{12}{\rm C}}:{^{16}{\rm O}}\sim 1:3:3$ by mass), this fast
expansion enables an $r$-process to occur in the shocked ejecta,
producing neutron-rich nuclei with $A>130$ through the actinides.

We discuss the conditions during the expansion of the shocked ejecta
in \S\ref{sec:conditions}. Nucleosynthesis calculations for two
different initial compositions are presented in
\S\ref{sec:nucleosynthesis} with and without 
neutrino reactions on free nucleons and $\alpha$-particles. We
discuss the implications of this new $r$-process model for
abundances in metal-poor stars and for general Galactic chemical
evolution in \S\ref{sec:gce}.

\section{Conditions During Expansion of Shocked Surface
Layers}\label{sec:conditions}

In the generic scenario of stellar core collapse, the inner core
becomes a proto-neutron star (PNS) and bounces on reaching
supra-nuclear density. This launches a shock, which propagates into
the still collapsing outer core and falters. According to the
neutrino-driven SN mechanism \citep{1985ApJ...295...14B}, the shock
is re-energized through heating of the material behind it by the
neutrinos emitted from the PNS. Using Nomoto's (1984, 1987) model
for a $1.38\,M_\odot$ O-Ne-Mg core, \citet{1988ApJ...334..909M}, and
most recently, \citet{2006A&A...450..345K} 
(see also \citealp{2007arXiv0706.3056J}) indeed obtained
neutrino-driven explosion although with very different final
explosion energies. For our purpose, the exact explosion energy does
not matter so long as the re-energized shock propagates through the
surface layers of the core with a sufficient speed. As only stars in
a very narrow range of $\sim 8$--$11\,M_\odot$ develop O-Ne-Mg
cores, we consider Nomoto's (1984, 1987) $1.38\,M_\odot$ core model,
which was evolved from the He core of a $\approx 9\,M_\odot$
progenitor, to be representative and adopt its quantitative
description of the pre-SN conditions. In addition, we use the
simulations of \citet{1988ApJ...334..909M} and
\citet{2006A&A...450..345K} based on this model as a guide to the
shock propagation.

At the onset of collapse, the core is surrounded by $\sim
0.1\,M_\odot$ of C and O, which in turn is surrounded by several
$10^{-6}\,M_\odot$ of He. We are interested in the outer C-O
layer that contains several $10^{-5}\,M_\odot$. This layer
is located at radii $r\approx 10^8$~cm and has a thickness
of $\approx 5\times10^6$~cm. The density $\rho$ within this region
changes from $10^6$ to $10^5$~g~cm$^{-3}$. The steepness of
this fall-off is measured by $w\equiv-d\ln\rho/d\ln r$, which is
61.7 at $\rho=3\times10^5$~g~cm$^{-3}$.
This can be understood from the condition of
hydrostatic equilibrium, $-(1/\rho)dP/dr=GM_C/r^2$, where $G$ is the
gravitational constant, $M_C=1.38\,M_\odot$ is the core mass, and
$P$ is the pressure provided by electrons and
nuclei with temperature $T\sim 6\times10^8$~K. 
The pre-SN density structure changes as a result of core collapse.
However, the local
free-fall timescale at $r\approx 10^8$~cm is
$\sim\sqrt{r^3/(GM_C)}\sim 0.1$~s, comparable to the time between
the onset of core collapse and shock arrival at this radius.
To zeroth order, we assume that the pre-SN density
structure in the region of interest remains unchanged before being
shocked.

The SN simulations mentioned above showed that
the shock speed in the region of interest is
$v_{\rm sh}\sim 10^{10}$~cm~s$^{-1}$.
For such $v_{\rm sh}$, the strong shock condition
$\rho v_{\rm sh}^2\gg P$ is satisfied and the energy density of the shocked
matter is expected to be dominated by the contributions from relativistic
particles (radiation and electron-positron pairs).
In this case, the density, velocity, and pressure of the shocked matter
(with the subscript ``$p$'' standing for ``post-shock'') are given by
(e.g., \citealp{1976ApJ...207..872C})
\begin{equation}
\rho_p=7\rho,\ v_p=\frac{6}{7}v_{\rm sh},\ P_p=\frac{6}{7}\rho
v_{\rm sh}^2, \label{eq:psh}
\end{equation}
respectively. Using Eq.~(\ref{eq:psh}) and
$P_p=(11/12)a_{\rm rad}T_p^4$ with $a_{\rm rad}$ being the radiation
constant, we obtain the post-shock temperature
\begin{equation}
T_p=1.05\times 10^{10}\rho_6^{1/4}v_{{\rm sh},10}^{1/2}\ {\rm K} \label{eq:tp}
\end{equation}
and the post-shock entropy in relativistic particles
\begin{equation}
S=\frac{11}{3}\frac{a_{\rm rad}T_p^3}{N_A\rho_p}=56.1
\frac{v_{{\rm sh},10}^{3/2}}{\rho_6^{1/4}}\ k\
{\rm nucleon}^{-1}, \label{eq:s}
\end{equation}
where $\rho_6$ is the pre-shock density in units of $10^6$~g~cm$^{-3}$,
$v_{{\rm sh},10}$ is the shock speed in units of $10^{10}$~cm~s$^{-1}$,
$N_A$ is Avogadro's number. The entropy $S$ is also a measure
of the energy density in relativistic particles relative to that in
nonrelativistic ones. Equation~(\ref{eq:s}) gives $S\gg 1$ in the region of
interest, so the corresponding energy density of the shocked matter
is indeed dominated by relativistic particles.

\citet{1999ApJ...510..379M} showed that
$v_{\rm sh}$ at a specific $r$ depends on the pre-shock
density at $r$ and the amount of matter entrained by the shock prior
to reaching $r$. As the region of interest is very thin and adds
very little to the matter already entrained by the shock, we expect
that the shock speed in this region takes the form
$v_{\rm sh}\propto\rho^{-0.19}$ \citep{1999ApJ...510..379M},
which gives
\begin{equation}
\frac{dv_{\rm sh}}{dr}=-0.19\left(\frac{d\ln\rho}{d\ln r}\right)
\frac{v_{\rm sh}}{r}=0.19w\left(\frac{v_{\rm sh}}{r}\right).\label{eq:vsh}
\end{equation}
We assume that matter moves with a constant velocity $v_p$ after
being shocked. Consider a mass element initially sandwiched between
radii $r$ and $r+\delta r$. When the matter at $r+\delta r$ is
shocked, the matter initially at $r$ has moved to $r+(v_p/v_{\rm
sh})\delta r= r+(6/7)\delta r$. The reduction of the mass element's
thickness from $\delta r$ to $\delta r/7$ results in the factor of 7
increase in its post-shock density (see Eq.~[\ref{eq:psh}]).
Subsequently, the matter at the inner and outer radii of the mass
element move with velocities of $v_p$ and $v_p+(dv_p/dr)\delta r$,
respectively. It can be shown that the density $\tilde{\rho}$ of
the shocked mass element evolves with time $t$ as:
\begin{equation}
\tilde{\rho}(t)=\frac{\rho_p}{[1+(t/\tau_1)]^2[1+(t/\tau_2)]},
\label{eq:rho}
\end{equation}
where $t=0$ is the time at which the mass element is shocked,
and
\begin{equation}
\tau_1=\frac{r}{v_p},\ \tau_2=\frac{1}{7(dv_p/dr)}
=\frac{\tau_1}{1.33w}. \label{eq:tau}
\end{equation}
In the denominator of
Eq.~(\ref{eq:rho}), the first factor accounts for the expansion of
the surface area of the mass element and the second for the
increase of its thickness. In relating $\tau_2$ to $\tau_1$ in
Eq.~(\ref{eq:tau}), we have used Eqs.~(\ref{eq:psh}) and (\ref{eq:vsh}).
To very good approximation, the expansion of the shocked
mass element is adiabatic. This then specifies the time evolution
of its temperature $\tilde{T}$ as
\begin{equation}
\tilde{T}(t)=\frac{T_p}{[1+(t/\tau_1)]^{2/3}[1+(t/\tau_2)]^{1/3}}.
\label{eq:temp}
\end{equation}

Considering the proximity of the region of interest to the PNS, we
also study how interaction with the intense neutrino flux affects
nucleosynthesis. Due to the short timescales $\tau_1\sim 10^{-2}$~s
and $\tau_2\sim 10^{-4}$~s
for the expansion of a shocked mass element, the neutrino luminosity
and energy spectra do not change during the expansion. We take the
luminosity to be $4\times 10^{52}$~erg~s$^{-1}$ per neutrino species,
$\langle E_\nu^2\rangle/\langle E_\nu\rangle=11$ and 16 MeV for
$\nu_e$ and $\bar\nu_e$, respectively, and $\langle
E_\nu\rangle=25$~MeV for $\nu_x=\nu_\mu$, $\bar\nu_\mu$, $\nu_\tau$,
$\bar\nu_\tau$, where $\langle E_\nu\rangle$ and $\langle E_\nu^2\rangle$
are the averages of the neutrino energy and its square, respectively, over
the relevant spectrum (the rates of interest for $\nu_e$ and $\bar\nu_e$
are essentially determined by the values of 
$\langle E_\nu^2\rangle/\langle E_\nu\rangle$ for these species and
insensitive to their detailed spectra, while the rates of interest for $\nu_x$
assume a black-body-like spectrum with a temperature of 8~MeV). We
include the reactions $\nu_e+n\to p+e^-$, $\bar\nu_e+p\to n+e^+$,
$\nu_x+{^4}{\rm He}\to{^3}{\rm He}+n$, and $\nu_x+{^4}{\rm
He}\to{^3}{\rm H}+p$, the rates of which are $0.253$, $0.250$,
$1.28\times 10^{-2}$, and $1.40\times 10^{-2}$~s$^{-1}$ per target,
respectively, at a radius of $10^8$~cm \citep{1996ApJ...471..331Q,
1990ApJ...356..272W}. For a mass element
initially at radius $r$, its radius $\tilde{r}$ evolves as
\begin{equation}
\tilde{r}(t)=r[1+(t/\tau_1)].
\end{equation}
All the neutrino reaction rates for this mass element scale as
$[\tilde{r}(t)]^{-2}$.

\section{Nucleosynthesis in Shocked Ejecta}\label{sec:nucleosynthesis}

We focus on a specific mass element initially at $r=10^8$~cm. The
pre-shock density structure at this position is characterized by
$\rho_6=0.3$ and $w=61.7$. We take $v_{{\rm sh},10}=1.5$ at
this position, which
gives $T_p=9.52\times 10^9$~K, $\tau_1=7.8\times 10^{-3}$~s, and
$\tau_2=9.48\times 10^{-5}$~s. The postshock density is
$\rho_p=2.1\times 10^6$~g~cm$^{-3}$, corresponding to an entropy
of $S=139$. We study the nucleosynthesis during the
expansion of the above mass element for two different initial
compositions: (1) $X(^{12}{\rm C})=X(^{16}{\rm O})=0.5$ corresponding
to $Y_e=0.5$ as in Nomoto's (1984, 1987) original model, and (2)
$X(^{13}{\rm C})=0.14$, $X(^{12}{\rm C})=X(^{16}{\rm O})=0.43$
corresponding to $Y_e=0.495$, which allows for possible production of
$^{13}$C due to mixing during the pre-SN evolution. Here,
for example, $X(^{12}{\rm C})$ is the mass fraction of $^{12}$C.
The above two cases are referred to as trajectories 1 and 2 when
neutrino reactions are not included and as trajectories 1$'$ and 2$'$
when these reactions are included.

To study the nucleosynthesis associated with trajectories 1 and 2,
we use the Clemson nuclear reaction network
\citep{2004ApJ...617L.131J}, which includes nuclear species from
neutrons and protons through the actinides.  Because the network flow
reached the top of the reaction network in some cases, we included
nuclear fission by allowing the most massive species in the network
($^{276}$U) to fission into two fragments (with $Z = 42$ and 50,
respectively).  This is a crude treatment that will require
improvement in follow-up calculations. Nevertheless, it illustrates
the effect of fission cycling.

\begin{figure}
\plotone{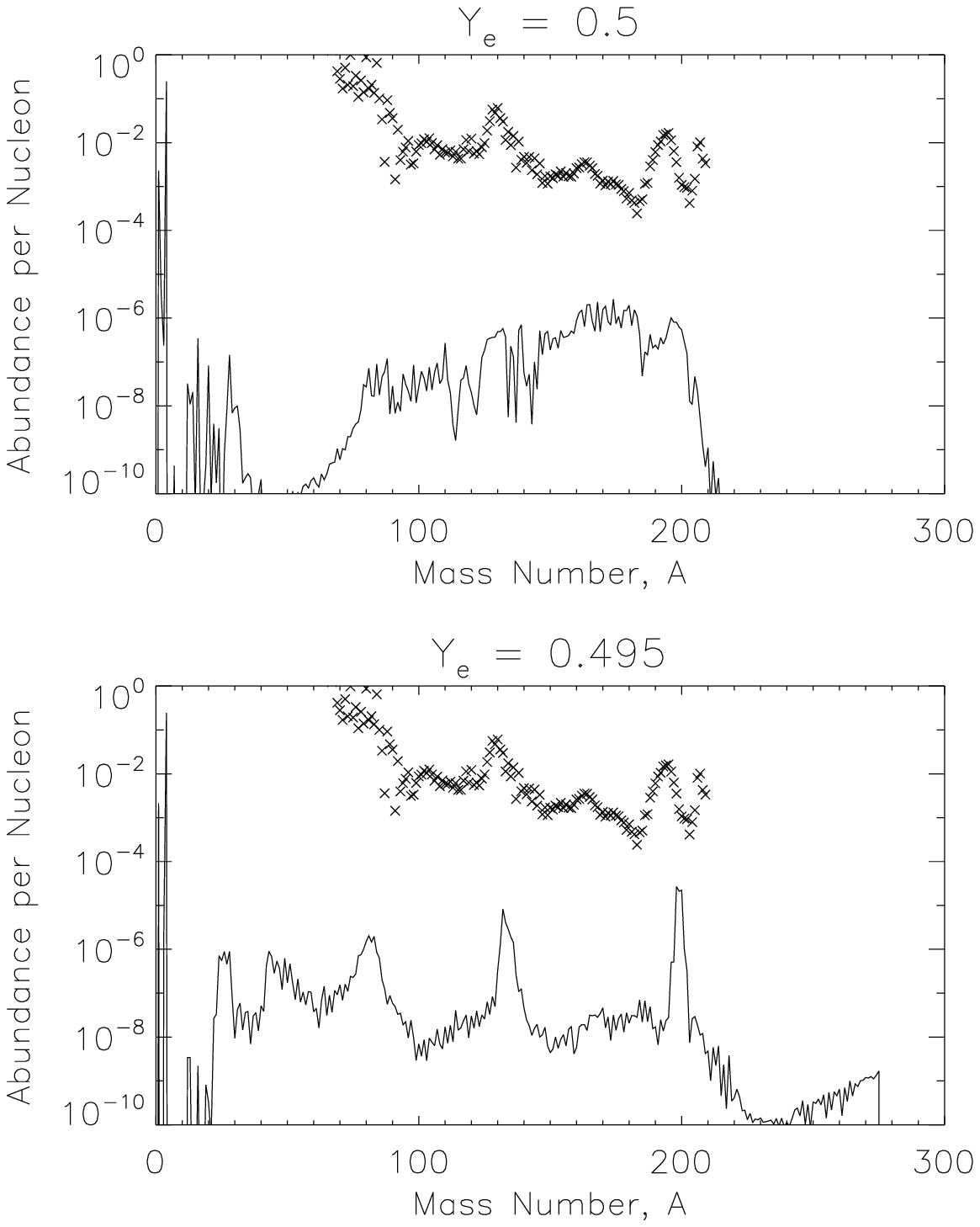}
\caption{Final abundances versus mass number for trajectories 1
    (top panel) and 2 (bottom panel).  The (arbitrarily scaled) solar $r$-process
   abundances \citep{1989RPPh...52..945K} are shown as $\times$'s for comparison.
   The final mass fractions resulting from both trajectories are
   $\approx 98$\% $\alpha$-particles and $\approx 2$\% heavy nuclei.}
\label{fig:fig1}
\end{figure}

Figure~\ref{fig:fig1} shows the final abundances for trajectories 1
(top panel) and 2 (bottom panel) as a function of $A$. Trajectory 1
(initial $Y_e = 0.5$) makes heavy, neutron-rich nuclei, although the
abundance pattern does not resemble the solar $r$-process abundance
distribution well. Trajectory 2 (with initial $Y_e = 0.495$) makes
heavy, neutron-rich nuclei, including actinides, and the resulting
abundance pattern shows three distinct peaks in rough agreement with
the solar $r$-process distribution. The peaks made by trajectory 2 lie 
several mass units above the corresponding solar $r$-process
abundance peaks. This is mainly due to the rather
extensive exposure to neutron capture for this trajectory: there are
still about 15 free neutrons per heavy nucleus at the end of the run.
The reason why trajectories 1 and 2 are able to produce heavy,
neutron-rich nuclei despite the zero to low initial neutron excess is 
the persistent nucleon--$\alpha$-particle disequilibrium discussed by
\citet{2002PhRvL..89w1101M}.  In these expansions, the initial
distribution of nuclei is quickly broken down into nuclear
statistical equilibrium, which is dominated by free neutrons
and protons. As the material expands and cools, the free nucleons
assemble into $\alpha$-particles and heavier nuclei.  Because of the
rapidness of the expansion, however, the free nucleons do not
assemble into heavier species (particularly $\alpha$-particles) as
quickly as equilibrium demands. This leaves a large excess of these
free nucleons, which push the heavy nuclei to very high mass.  For
example, for trajectory 2, at $T_9 \approx 4$, the abundance
distribution is dominated by free neutrons and protons, $\alpha$-particles, 
and nuclei with mass number $A \approx 140$. Specifically, the
mass fractions of free neutrons and protons, $\alpha$-particles,
and heavy nuclei at this temperature are $\approx 1.3$\%, 0.3\%, 97.6\%, 
and 0.6\%, respectively. The heavy
nuclei serve as the seeds for the subsequent $r$-process
nucleosynthesis.

We expect that the final abundance distribution will depend on the
details of the trajectories of the mass elements at late times.
Indeed, we note that for trajectory 2, expansion occurs so rapidly
that reaction freezeout occurs because the density drops to such a
low value that neutron-capture timescales become long, not because
the neutrons are all consumed. The result is that there are still
about 15 free neutrons per heavy nucleus at the end of the run
(the final mass fractions of free neutrons, 
$\alpha$-particles, and heavy nuclei are $\approx 0.2$\%,
98\%, and 2\%, respectively). We
confirmed this by running a calculation identical to trajectory 2
except that the expansion slowed at late times and allowed the
neutrons to be all consumed.  The resulting pattern was broadly
similar to that in the lower panel of Fig.~\ref{fig:fig1}, but the
contrast between the peaks and valleys was smaller.

We also followed trajectories 1$'$ and  2$'$, which were identical to
trajectories 1 and 2 except that neutrino reactions were included.  The
neutrino interactions had negligible effects on the abundance
patterns.  This is because expansion carried the material away so
rapidly that the total number of neutrino interactions per nucleon or
nucleus was $\ll 1$ during the expansion.

\section{Discussion and Conclusions}\label{sec:gce}

We have proposed a new model of $r$-process nucleosynthesis in
O-Ne-Mg-core collapse SNe. Unlike previous models based on
assumed extremely neutron-rich ejecta (e.g.,
\citealp{1998ApJ...493L.101W,2003ApJ...593..968W}),
our model relies on the fast expansion of the shocked matter in
the weakly neutron-rich surface layers of an O-Ne-Mg core to
achieve a disequilibrium between free nucleons, $\alpha$-particles,
and heavier nuclei. As shown by \citet{2002PhRvL..89w1101M},
the significant presence of free nucleons
in disequilibrium facilitates the production of seed nuclei with
$A\sim 140$ during the expansion. For matter
with an initial $Y_e=0.495$, the neutron excess is sufficient for
an $r$-process with fission cycling to occur, producing dominantly
the nuclei with $A>130$ through the
actinides. The key to the fast expansion of
the shocked matter is the steep density gradient in the surface layers
of an O-Ne-Mg core, which enables the inner and outer surfaces of
a mass element to expand with significantly different velocities, thus
making its density drop much faster.

O-Ne-Mg-core collapse SNe were also proposed as
the source for the heavy $r$-process elements ($r$-elements)
with $A>130$ (Ba and higher atomic numbers) based on
considerations of Galactic chemical evolution
(e.g., \citealp{1992ApJ...391..719M,1999ApJ...511L..33I}).
A strong argument for this proposal was presented using
observations of metal-poor stars
\citep{2002ApJ...567..515Q,2003ApJ...588.1099Q}.
In addition, the key issues regarding the sources for the
$r$-elements based on observations and basic understanding of
stellar models were summarized in \citet{2007PhR...442..237Q}.
Data on the metal-poor stars CS~31082--001 \citep{2002A&A...387..560H},
HD~115444, and HD~122563 \citep{2000ApJ...530..783W}  show that
their abundances of the heavy $r$-elements differ by a factor up to
$\sim 10^2$. In contrast, these stars have essentially the same
abundances of the elements between O and Ge (e.g.,
[Fe/H]~$\equiv\log({\rm Fe/H})-\log({\rm Fe/H})_\odot\sim -3$).
Furthermore, when CS~22892--052 ([Fe/H]~$=-3.1$,
\citealp{2003ApJ...591..936S})
is compared with HD~221170 ([Fe/H]~$=-2.2$, \citealp{2006ApJ...645..613I})
and CS~31082--001 ([Fe/H]~$=-2.9$) with BD~$+17^\circ 3248$
([Fe/H]~$=-2.1$, \citealp{2002ApJ...572..861C}), data show that
the stars in either pair have nearly the same abundances of heavy
$r$-elements but the abundances of the elements between O and Ge
differ by a factor of $\sim 8$ and 6 for the former and latter pair, respectively.
These results appear to require that the production of the heavy $r$-elements
be decoupled from that of the elements between O and Ge
\citep{2002ApJ...567..515Q,2003ApJ...588.1099Q,2007PhR...442..237Q}.
As Fe-core collapse SNe from progenitors of $>11\,M_\odot$ are the major
source for the latter group of elements at low metallicities, this strongly
suggests that such SNe are not the source for the heavy $r$-elements.
The elements between O and Ge are produced between the core and
the H envelope by explosive burning during a core-collapse SN or by
hydrostatic burning during its pre-SN evolution.
Models of O-Ne-Mg-core collapse SNe show that
the total amount of material ejected from between the core and the H
envelope is only $\sim 0.01$--$0.04\,M_\odot$
\citep{1988ApJ...334..909M,2006A&A...450..345K}, much smaller than
the $\sim 1\,M_\odot$ for Fe-core collapse SNe. Thus, O-Ne-Mg-core
collapse SNe contribute very little to the elements between O and Ge.
The decoupling between these elements and the heavy $r$-elements
can then be explained by attributing the heavy $r$-elements to such SNe
as argued in 
\citet{2002ApJ...567..515Q,2003ApJ...588.1099Q,2007PhR...442..237Q}. 
This attribution is supported by the $r$-process model proposed here.

The collapse of O-Ne-Mg cores may comprise up to 
$\sim 20\%$ of all core-collapse events and may thus occur roughly once a
century (e.g., \citealp{2007arXiv0705.4643P}).  Assume that this frequency
of occurrence is associated with a reference gas of $\sim 10^{10}\,M_\odot$
and that the O-Ne-Mg-core collapse SNe over the Galactic history of 
$\sim 10^{10}$~yr enriched this gas with
a solar mass fraction of $\sim 4\times 10^{-8}$ of heavy $r$-elements.
We find that each SN
must produce $\sim 4\times10^{-6}\,M_\odot$ of heavy $r$-elements.
The total mass fraction of heavy $r$-elements produced during the
expansion associated with trajectory 2 is $\sim 0.02$ and a total
$\sim 3\times10^{-5}\,M_\odot$ of ejecta may experience such
expansion in Nomoto's (1984, 1987) O-Ne-Mg core model. This assumes
an initial composition with $Y_e=0.495$ and would give $\sim 6\times
10^{-7}\,M_\odot$ of heavy $r$-elements per SN. A slightly smaller
$Y_e=0.49$ would double the initial neutron excess and the final
$r$-process production, giving $\sim10^{-6}\,M_\odot$ of heavy
$r$-elements. Thus, we consider the model proposed here has the
potential to account for the Galactic inventory of such elements. To
further test this model, we urge two lines of important studies: (1)
calculating the evolution of $\sim 8$--$11\,M_\odot$ stars to
determine the pre-SN conditions of O-Ne-Mg cores, especially the
neutron excess and density structure of the surface layers; and (2)
simulating the collapse of such cores and the subsequent shock
propagation to determine the conditions of the shocked surface
layers. As these layers contain very little mass, simulations with
extremely fine mass resolutions are required to demonstrate the fast
expansion of shocked ejecta that is proposed here as the key to the
production of heavy $r$-elements. The pre-SN abundance of
$^{13}$C in the surface layers also requires investigation as this
appears critical to the total yields of heavy $r$-elements.

\acknowledgments

We thank Ken Nomoto for providing his O-Ne-Mg core model and
for valuable discussion. We also thank Jerry Wasserburg and an
anonymous referee for helpful comments.
This work was supported in part by DOE grant DE-FG02-87ER40328 and
NASA Cosmochemistry Program grant NNX07AJ04G.

\clearpage

\begin{thebibliography}{31}
\expandafter\ifx\csname natexlab\endcsname\relax\def\natexlab#1{#1}\fi

\bibitem[{{Bethe} \& {Wilson}(1985)}]{1985ApJ...295...14B}
{Bethe}, H.~A. \& {Wilson}, J.~R. 1985, \apj, 295, 14

\bibitem[{{Chevalier}(1976)}]{1976ApJ...207..872C}
{Chevalier}, R.~A. 1976, \apj, 207, 872

\bibitem[{{Cowan} {et~al.}(2002){Cowan}, {Sneden}, {Burles}, {Ivans}, {Beers},
  {Truran}, {Lawler}, {Primas}, {Fuller}, {Pfeiffer}, \&
  {Kratz}}]{2002ApJ...572..861C}
{Cowan}, J.~J., {Sneden}, C., {Burles}, S., {Ivans}, I.~I., {Beers}, T.~C.,
  {Truran}, J.~W., {Lawler}, J.~E., {Primas}, F., {Fuller}, G.~M., {Pfeiffer},
  B., \& {Kratz}, K.-L. 2002, \apj, 572, 861

\bibitem[{{Garcia-Berro} \& {Iben}(1994)}]{1994ApJ...434..306G}
{Garcia-Berro}, E. \& {Iben}, I. 1994, \apj, 434, 306

\bibitem[{{Garcia-Berro} {et~al.}(1997){Garcia-Berro}, {Ritossa}, \&
  {Iben}}]{1997ApJ...485..765G}
{Garcia-Berro}, E., {Ritossa}, C., \& {Iben}, I.~J. 1997, \apj, 485, 765

\bibitem[{{Hill} {et~al.}(2002){Hill}, {Plez}, {Cayrel}, {Beers},
  {Nordstr{\"o}m}, {Andersen}, {Spite}, {Spite}, {Barbuy}, {Bonifacio},
  {Depagne}, {Fran{\c c}ois}, \& {Primas}}]{2002A&A...387..560H}
{Hill}, V., {Plez}, B., {Cayrel}, R., {Beers}, T.~C., {Nordstr{\"o}m}, B.,
  {Andersen}, J., {Spite}, M., {Spite}, F., {Barbuy}, B., {Bonifacio}, P.,
  {Depagne}, E., {Fran{\c c}ois}, P., \& {Primas}, F. 2002, \aap, 387, 560

\bibitem[{{Iben} {et~al.}(1997){Iben}, {Ritossa}, \&
  {Garcia-Berro}}]{1997ApJ...489..772I}
{Iben}, I.~J., {Ritossa}, C., \& {Garcia-Berro}, E. 1997, \apj, 489, 772

\bibitem[{{Ishimaru} \& {Wanajo}(1999)}]{1999ApJ...511L..33I}
{Ishimaru}, Y. \& {Wanajo}, S. 1999, \apjl, 511, L33

\bibitem[{{Ivans} {et~al.}(2006){Ivans}, {Simmerer}, {Sneden}, {Lawler},
  {Cowan}, {Gallino}, \& {Bisterzo}}]{2006ApJ...645..613I}
{Ivans}, I.~I., {Simmerer}, J., {Sneden}, C., {Lawler}, J.~E., {Cowan}, J.~J.,
  {Gallino}, R., \& {Bisterzo}, S. 2006, \apj, 645, 613

\bibitem[{{Janka} {et~al.}(2007){Janka}, {Marek}, \&
  {Kitaura}}]{2007arXiv0706.3056J}
{Janka}, H.-T., {Marek}, A., \& {Kitaura}, F.-S. 2007, ArXiv e-prints, 706

\bibitem[{{Jordan} \& {Meyer}(2004)}]{2004ApJ...617L.131J}
{Jordan}, IV, G.~C. \& {Meyer}, B.~S. 2004, \apjl, 617, L131

\bibitem[{{Kappeler} {et~al.}(1989){Kappeler}, {Beer}, \&
  {Wisshak}}]{1989RPPh...52..945K}
{Kappeler}, F., {Beer}, H., \& {Wisshak}, K. 1989, Reports of Progress in
  Physics, 52, 945

\bibitem[{{Kitaura} {et~al.}(2006){Kitaura}, {Janka}, \&
  {Hillebrandt}}]{2006A&A...450..345K}
{Kitaura}, F.~S., {Janka}, H.-T., \& {Hillebrandt}, W. 2006, \aap, 450, 345

\bibitem[{{Mathews} {et~al.}(1992){Mathews}, {Bazan}, \&
  {Cowan}}]{1992ApJ...391..719M}
{Mathews}, G.~J., {Bazan}, G., \& {Cowan}, J.~J. 1992, \apj, 391, 719

\bibitem[{{Matzner} \& {McKee}(1999)}]{1999ApJ...510..379M}
{Matzner}, C.~D. \& {McKee}, C.~F. 1999, \apj, 510, 379

\bibitem[{{Mayle} \& {Wilson}(1988)}]{1988ApJ...334..909M}
{Mayle}, R. \& {Wilson}, J.~R. 1988, \apj, 334, 909

\bibitem[{{Meyer}(2002)}]{2002PhRvL..89w1101M}
{Meyer}, B.~S. 2002, Physical Review Letters, 89, 231101

\bibitem[{{Nomoto}(1984)}]{1984ApJ...277..791N}
{Nomoto}, K. 1984, \apj, 277, 791

\bibitem[{{Nomoto}(1987)}]{1987ApJ...322..206N}
---. 1987, \apj, 322, 206

\bibitem[{{Poelarends} {et~al.}(2007){Poelarends}, {Herwig}, {Langer}, \&
  {Heger}}]{2007arXiv0705.4643P}
{Poelarends}, A.~J.~T., {Herwig}, F., {Langer}, N., \& {Heger}, A. 2007, ArXiv
  e-prints, 705

\bibitem[{{Qian} \& {Wasserburg}(2002)}]{2002ApJ...567..515Q}
{Qian}, Y.-Z. \& {Wasserburg}, G.~J. 2002, \apj, 567, 515

\bibitem[{{Qian} \& {Wasserburg}(2003)}]{2003ApJ...588.1099Q}
---. 2003, \apj, 588, 1099

\bibitem[{{Qian} \& {Wasserburg}(2007)}]{2007PhR...442..237Q}
---. 2007, \physrep, 442, 237

\bibitem[{{Qian} \& {Woosley}(1996)}]{1996ApJ...471..331Q}
{Qian}, Y.-Z. \& {Woosley}, S.~E. 1996, \apj, 471, 331

\bibitem[{{Ritossa} {et~al.}(1996){Ritossa}, {Garcia-Berro}, \&
  {Iben}}]{1996ApJ...460..489R}
{Ritossa}, C., {Garcia-Berro}, E., \& {Iben}, I.~J. 1996, \apj, 460, 489

\bibitem[{{Ritossa} {et~al.}(1999){Ritossa}, {Garc{\'{\i}}a-Berro}, \&
  {Iben}}]{1999ApJ...515..381R}
{Ritossa}, C., {Garc{\'{\i}}a-Berro}, E., \& {Iben}, I.~J. 1999, \apj, 515, 381

\bibitem[{{Sneden} {et~al.}(2003){Sneden}, {Cowan}, {Lawler}, {Ivans},
  {Burles}, {Beers}, {Primas}, {Hill}, {Truran}, {Fuller}, {Pfeiffer}, \&
  {Kratz}}]{2003ApJ...591..936S}
{Sneden}, C., {Cowan}, J.~J., {Lawler}, J.~E., {Ivans}, I.~I., {Burles}, S.,
  {Beers}, T.~C., {Primas}, F., {Hill}, V., {Truran}, J.~W., {Fuller}, G.~M.,
  {Pfeiffer}, B., \& {Kratz}, K.-L. 2003, \apj, 591, 936

\bibitem[{{Wanajo} {et~al.}(2003){Wanajo}, {Tamamura}, {Itoh}, {Nomoto},
  {Ishimaru}, {Beers}, \& {Nozawa}}]{2003ApJ...593..968W}
{Wanajo}, S., {Tamamura}, M., {Itoh}, N., {Nomoto}, K., {Ishimaru}, Y.,
  {Beers}, T.~C., \& {Nozawa}, S. 2003, \apj, 593, 968

\bibitem[{{Westin} {et~al.}(2000){Westin}, {Sneden}, {Gustafsson}, \&
  {Cowan}}]{2000ApJ...530..783W}
{Westin}, J., {Sneden}, C., {Gustafsson}, B., \& {Cowan}, J.~J. 2000, \apj,
  530, 783

\bibitem[{{Wheeler} {et~al.}(1998){Wheeler}, {Cowan}, \&
  {Hillebrandt}}]{1998ApJ...493L.101W}
{Wheeler}, J.~C., {Cowan}, J.~J., \& {Hillebrandt}, W. 1998, \apjl, 493, L101+

\bibitem[{{Woosley} {et~al.}(1990){Woosley}, {Hartmann}, {Hoffman}, \&
  {Haxton}}]{1990ApJ...356..272W}
{Woosley}, S.~E., {Hartmann}, D.~H., {Hoffman}, R.~D., \& {Haxton}, W.~C. 1990,
  \apj, 356, 272

\end{thebibliography}
\end{document}